\documentclass{aip-cp}

\usepackage[numbers]{natbib}
\usepackage{rotating}
\usepackage{graphicx}

\begin{document}

\title{Physics Opportunities with a Secondary $K_L^0$ Beam at JLab$^{\footnote {Talk at  XVI International Conference on Hadron Spectroscopy, September 2015, Newport News, Virginia, USA }}$}

\author[aff1]{Moskov Amaryan\corref{cor1}}

\affil[aff1]{Old Dominion University, Norfolk, VA, USA.}

\corresp[cor1]{mamaryan@odu.edu}

\maketitle

\begin{abstract}
Following a Letter of Intent submitted to PAC43 at JLab in this talk we discuss the possibility to create a secondary $K_L^0$ beam in Hall-D to be used with GlueX detector for spectroscopy of excited hyperons.

\end{abstract}

\section{INTRODUCTION}

Our current understanding of  strong interactions is embedded in Quantum Chromodynamics (QCD). However, QCD being a basic theory, extremely successful in explaining the plethora of experimental data in the perturbative regime, faces significant challenges to describe the properties of hadrons in non-perturbative regime.  Constituent Quark Model (CQM) is surprisingly successful in explaining spectra of  hadrons,  especially in the ground state; however, CQM appears to be too naive to describe properties of excited states. It is natural that excited states are not simply explained with spatial excitations of constituent quarks, but it is an effective representation revealing complicated interactions of quarks and gluons inside. Hadron spectroscopy aims to provide a comprehensive description of hadron structure based on quark and gluon degrees of freedom. Despite many successes in observing hundreds of meson and baryon states experimentally we haven't succeeded to either observe or rule out existence of glueballs, hybrids and multi quark systems; although it is tempting to explain recently observed  X, Y, Z ~\cite{bib1} states as first evidences of tetraquarks as well as recently observed heavy baryon states at LHCb ~\cite{bib2} as charmed pentaquarks.

An extensive experimental program is developed to search for hybrids in the GlueX experiment at JLab. Over the last decade, significant progress in our understanding of baryons made of light $(u,d)$ quarks have been made in CLAS at JLab. However, systematic studies of excited hyperons are very much lacking with only decades old very scarce data filling the world database in many channels. In this experiment we propose to fill this gap and study spectra of excited hyperons using the modern CEBAF facility with the aim to use proposed secondary $K^0_L$ beam with physics target of the GlueX experiment in Hall~D. The goal is to study $K_L-p$ and $K_L-d$ interactions and do the baryon spectroscopy for the strange baryon sector.

Unlike in the cases with pion or photon beams, kaon beams are crucial to provide the data needed to identify and characterize the properties of hyperon resonances. 

Our current experimental knowledge of strange resonances is far worse than our knowledge of $N$ and $\Delta$ resonances; however, within the quark model, they are no less fundamental. Clearly there is a need to learn about baryon resonances in the ``strange sector" to have a complete understanding of three-quark bound states.

The masses and widths of the lowest mass baryons were determined with kaon-beam experiments in the 1970s  ~\cite{bib1}. First determination of pole positions, for instance for $\Lambda(1520)$, were obtained only recently from analysis of Hall~A measurement at JLab ~\cite{bib3}. An intense kaon beam would open a window of opportunity not only to locate missing resonances, but also to establish properties including decay channels systematically for higher excited states.

A comprehensive review of physics opportunities with meson beams is presented in a recent paper ~\cite{bib4}. Importance of baryon spectroscopy in strangeness sector was  discussed in 
Ref.~\cite{bib5}.

\section{Reactions that could be studied with $K_L^0$ beam}
\subsubsection{ Elastic and charge-exchange reactions}

\begin{eqnarray}
	K_L^0p\to K_S^0p\\
	K_L^0p\to K^+n
\end{eqnarray}

\subsubsection{Two-body reactions producing $S=-1$ hyperons}

\begin{eqnarray}
	K_L^0p\to \pi^+\Lambda\\
	K_L^0p\to \pi^+\Sigma^0
\end{eqnarray}

\subsubsection{Three-body reactions producing $S=-1$ hyperons}

\begin{eqnarray}
 	K_L^0p\to \pi^+\pi^0\Lambda \\
 	K_L^0p\to \pi^+\pi^0\Sigma^0 \\
 	K_L^0p\to \pi^0\pi^0\Sigma^+  \\
 	K_L^0p\to \pi^+\pi^- \Sigma^+  \\
 	K_L^0p\to \pi^+\pi^- \Sigma^- 
\end{eqnarray}

\subsubsection{Two- and three-body reactions producing $S=-2$ hyperons} 

\begin{eqnarray}
	K_L^0p\to K^+\Xi^0 \\
	K_L^0p\to \pi^+K^+\Xi^-\\
	K_L^0p\to K^+\Xi^{0*} \\
	K_L^0p\to \pi^+K^+\Xi^{-*}
\end{eqnarray}

\subsubsection{Three-body reactions producing $S=-3$ hyperons}
\begin{eqnarray}
	K_L^0p\to K^+K^+\Omega^-\\
	K_L^0p\to K^+K^+\Omega^{-*}
\end{eqnarray}

Reactions 10-15 will be discussed in more detail below.
\section{The $K_L^0$ BEAM in HALL D}

In this chapter we describe photo-production of secondary $K_L^0$ beam in Hall~D. 

\begin{figure}[htb!]
\centerline{\includegraphics[width=350pt]{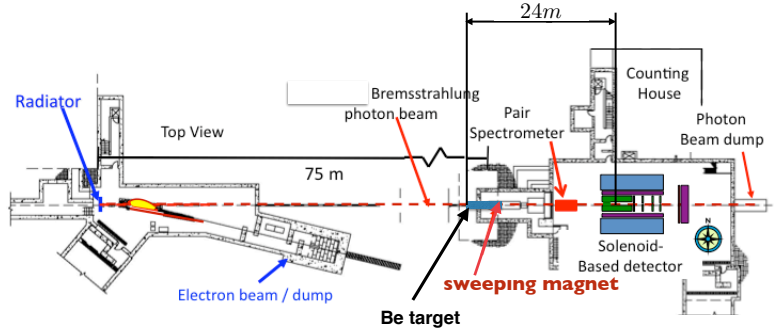}}
 \caption{ Schematic view of Hall~D beamline. See a text for explanation.
		 \label{fig:setup}}
\end{figure}

At  the first stage, $E_e=12$~GeV electrons produced at CEBAF will scatter in a radiator in the tagger vault,  generating  intensive beam of bremsstrahlung photons. At the second stage, bremsstrahlung photons interact with Be target placed on a distance 16~m upstream of liquid hydrogen ($LH_2$) target  of GlueX experiment in Hall D producing $K_L^0$ beam. To stop photons a 30 radiation length  lead absorber will be installed in the beamline followed by sweeping magnet to deflect the flow of charged particles. The flux of $K_L$ on ($LH_2$) target of GlueX experiment in Hall~D will be measured with pair spectrometer upstream the target. Momenta of $K_L$ particles will be measured using the time-of-flight between RF signal of CEBAF and start counters surrounding  $LH_2$ target. Schematic view of beamline is presented in Fig.\ref{fig:setup}. The bremsstrahlung photons, created by electrons at a distance about 75~m upstream, hit the Be target and produce $K_L^0$ mesons along with neutrons and charged particles. The lead absorber of $\sim$30 radiation length is installed to absorb photons  exiting Be target. The sweeping magnet deflects any remaining charged particles (leptons or hadrons) remaining after the absorber. The pair spectrometer will monitor the flux  of $K_L^0$ through the decay rate of kaons at given distance about 10~m from Be target.


Here we outline experimental conditions and simulated flux of $K_L^0$ based on GEANT4 and known cross sections of underlying subprocesses ~\cite{bib6,bib7,bib8}.

\begin{itemize}
\item An electron beam with energy $E_e=12$~GeV and current $I_e = 
	5~\mu A$ (maximum possible, limited by the Hall D beam dump).
\item A thickness of radiator 5~$\%$ radiation length.
\item Primary Be target with $R=4$~cm, $L=40$~cm.
\item $LH_2$ target with $R=2$~cm, $L=30$~cm.
\item Distance between Be and $LH_2$ targets $24$~m.
\end{itemize}

The expected flux of $K_L^0$ mesons integrated in the range of momenta $P=0.3-10 GeV/c$  will be $\approx 2\times 10^3$~$K_L^0/sec$ on the physics target of the GlueX setup.
\vskip 0.5cm
In a more aggressive scenario with 

\begin{itemize}
\item A thickness of radiator 10$\%$.
\item Be target with a length $L=60$~cm.
\item $LH_2$ target with $R=3$~cm.
\end{itemize}

The expected flux of $K_L^0$ mesons integrated over the same momentum range  will increase to $\approx 10^4 K_L^0/sec$.

In addition to these requirements it will require lower repetition rate of electron beam with $\sim 40$~ns spacing between bunches to have enough time to measure time-of-flight of the beam momenta and  to avoid an overlap of events produced from alternating pulses. Lower repetition rate was already successfully used by G0 experiment in Hall C at JLab ~\cite{bib9}.

The radiation length of the radiator needs further studies in order to estimate the level of radiation and required shielding in the tagger region. During this experiment all photon beam tagging detector systems and electronics will be removed.

The final flux of $K_L^0$ is presented with 10$\%$ radiator, corresponding to maximal rate .

\begin{figure}[htb!]
\includegraphics[width=2.0in]{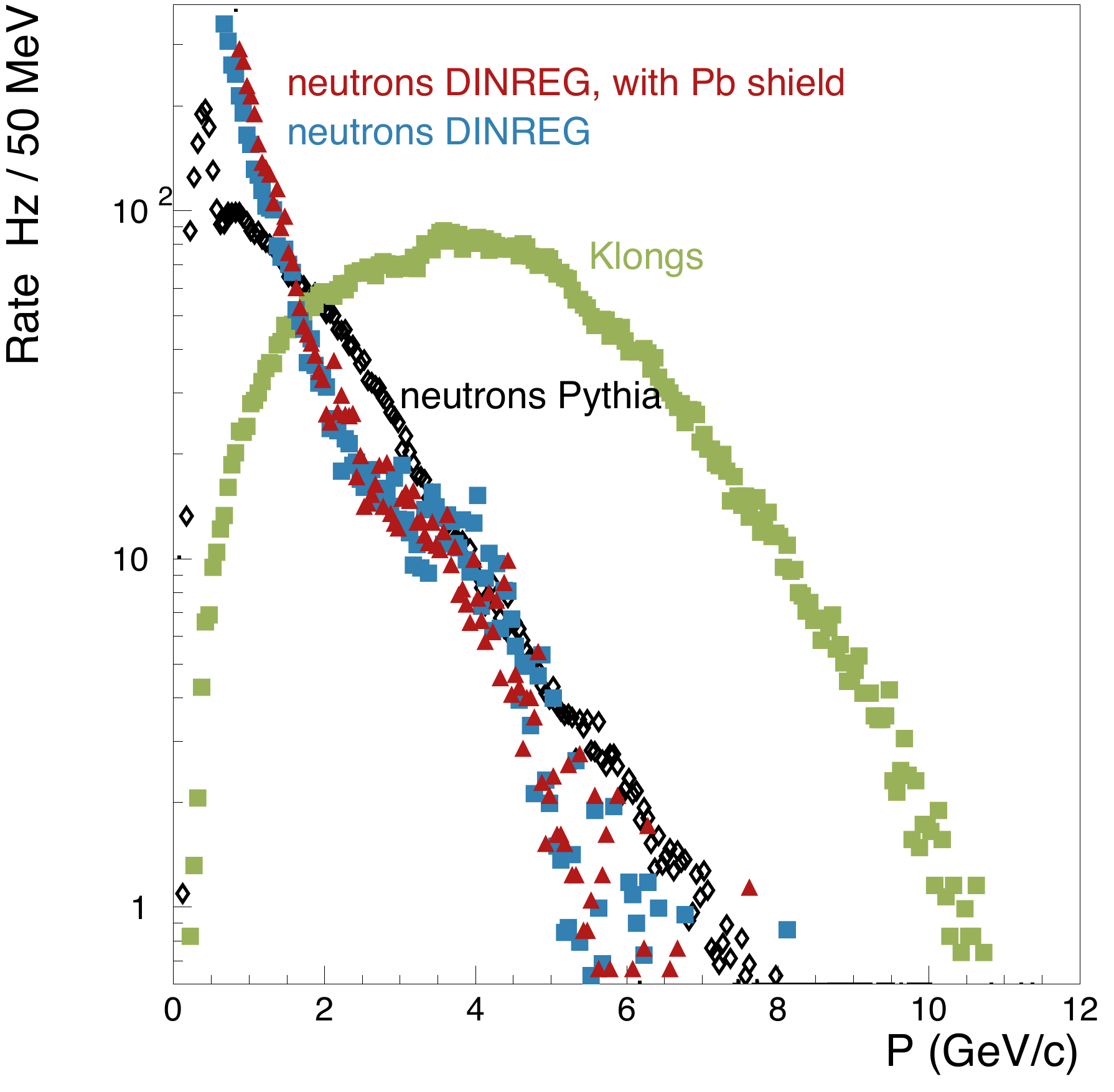}
 \caption{\baselineskip 13pt The rate of neutrons (open symbols) and 	$K_L^0$ (full squares) on $LH_2$ target of Hall~D as a function 	of their momenta simulated with different MC generators with 
	$10^4K^0_L$/sec.
 		 \label{fig:ratio}}
\end{figure}

In the production of a beam of neutral kaons, an important factor is the rate of neutrons as background. As it is well known, the ratio $R=N_n/N_{K_L^0}$ is on the order $10^3$ from primary proton beams~\cite{bib10}, the same ratio with primary electromagnetic interactions is much lower. This is illustrated in Fig.\ref{fig:ratio}, which  presents the rate of kaons and neutrons as a function of the momentum, which resembles similar behavior as it was measured at SLAC~\cite{bib11}.

\section{EXPECTED RATES}

\begin{figure}[htb!]
\includegraphics[width=3.0in, height=2.2in, trim=0 0 0 0cm]{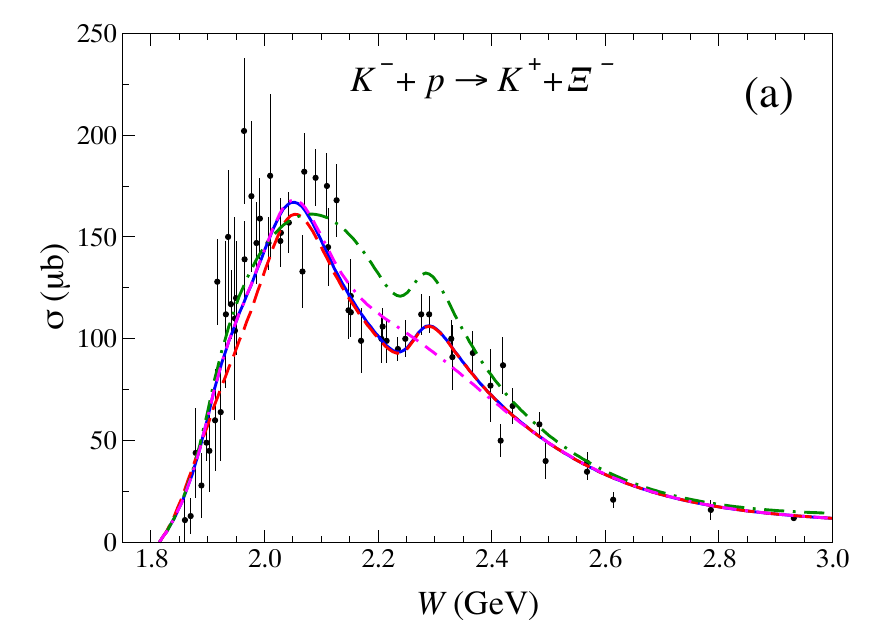}
\includegraphics[width=3.0in, height=2.50in, trim=0 3.0cm 1cm 0cm]{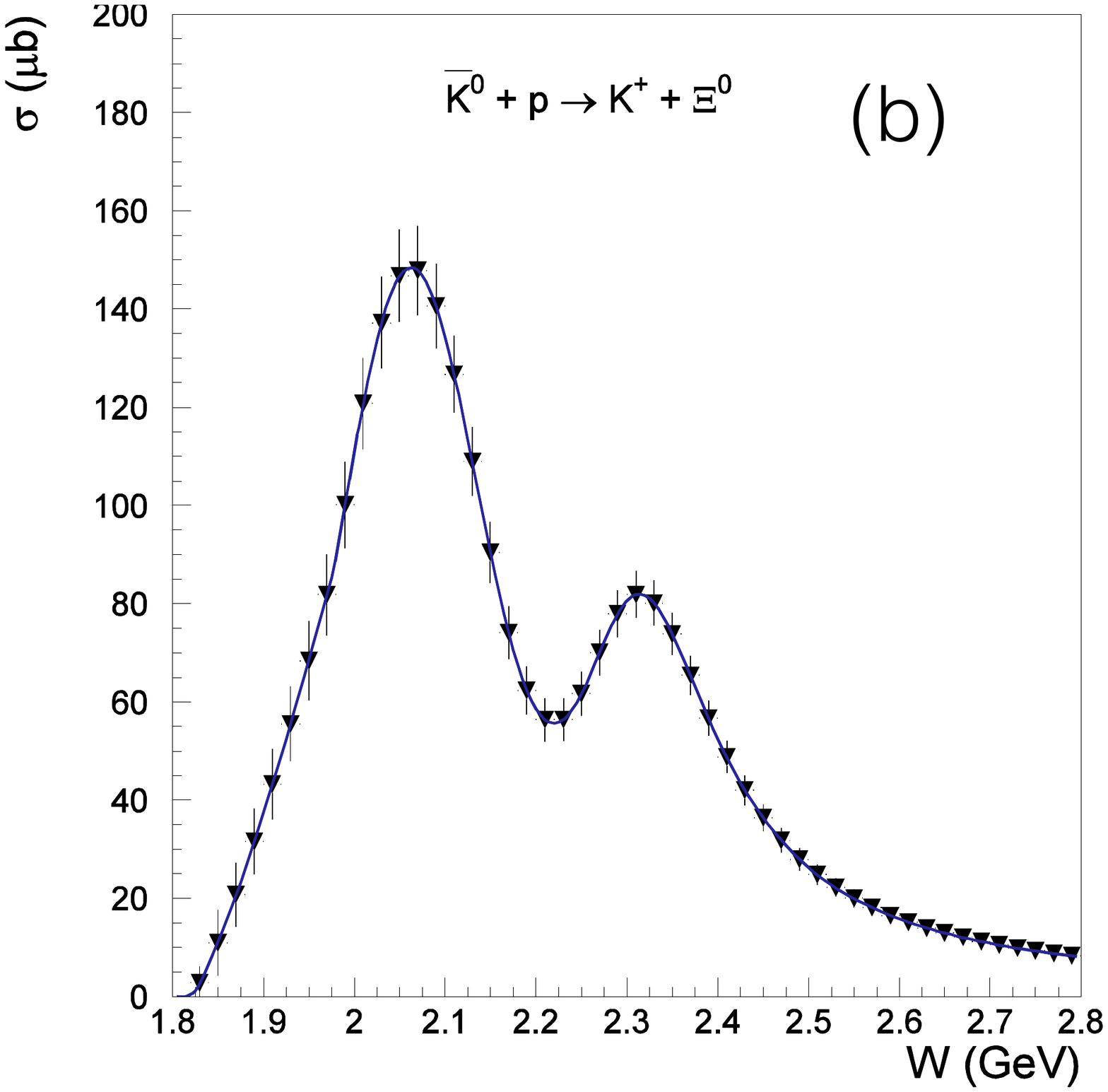}
 \caption{\baselineskip 13pt (Color online). a) Cross section for existing 	world data on $K^-+p\to K^+\Xi^-$ reaction with model predictions from ~\cite{bib12}; b) expected statistical precision for the 	reaction $K_L^0+p\to K^+\Xi^0$ in 10 days of running with a beam intensity $2\times 10^3 K_L$/sec overlaid on theoretical 
	prediction~\cite{bib12}. 
	 		 \label{fig:xi_prod}}
\end{figure}

In this section we discuss expected rates of events for some selected reactions. The production of $\Xi$ hyperons has been measured only with charged kaons with very low statistical precision and never with primary $K_L^0$ beam. In Fig.\ref{fig:xi_prod} left panel shows existing data for the octet ground state $\Xi$'s with theoretical model predictions for $W$ (the reaction center of mass energy)  distribution. On the right panel, the same model prediction ~\cite{bib12} is presented with expected experimental points and statistical error for 10 days of running with our proposed setup with a beam intensity $2\times 10^3 K_L$/sec is presented using missing mass of $K^+$ in the reaction $K_L^0+p\to K^+\Xi^0$ without detection of any decay products of $\Xi^0$.

\begin{figure}[htb!]
\includegraphics[width=4.0in, height=3.0in ]{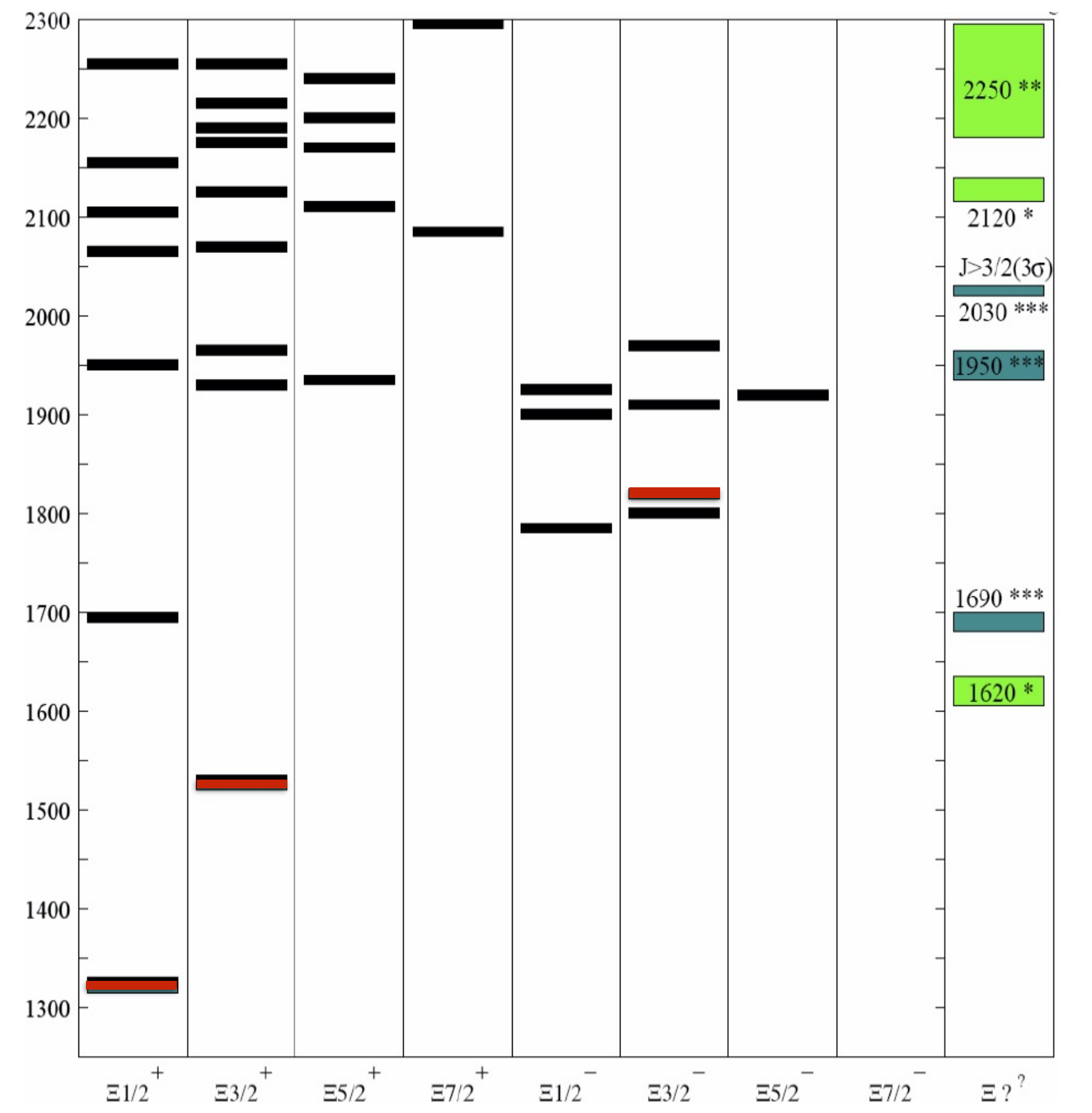}
 \caption{\baselineskip 13pt (Color online). Black bars: Predicted $\Xi$ spectrum based 	on the quark model calculation~\protect\cite{bib13}. Colored 	bars: Observed states. The two ground octet and decuplet states 	together with $\Xi(1820)$ in the column $J^P=3/2^-$ are shown in 	red color. Other observed states with unidentified spin-parity  	are plotted in the rightest column.
 		 \label{fig:table1}}
\end{figure}

The physics of excited hyperons is not well explored, remaining essentially at the pioneering stages of '70s-'80s. This is especially true for $\Xi^\ast(S=-2)$  and $\Omega^\ast(S=-3)$ hyperons.
For example, the $SU(3)$ flavor symmetry allows as many $S=-2$ baryon resonances, as there are $N$ and $\Delta$ resonances combined 
($\approx 27$); however, until now only three [ground state $\Xi(1382)1/2^+$, $\Xi(1538)3/2^+$, and $\Xi(1820)3/2^-$] have their quantum numbers assigned and few more states have been observed~\cite{bib1}. The status of $\Xi$ baryons is summarized In a table presented in 
Fig.\ref{fig:table1} together with quark model predicted states ~\cite{bib13}.

Historically the $\Xi^\ast$ states were intensively searched for mainly in bubble chamber experiments using the $K^-p$ reaction in '60s-'70s. The cross section was estimated to be on the order of 1-10~$\mu b$ at the beam momenta up to ~10~GeV/c. In '80s-'90s, the mass or width of ground and some of excited states were measured with a spectrometer in the CERN hyperon beam experiment. Few experiments have studied cascade baryons with the missing mass technique. In 1983, the production of $\Xi^*$ resonances up to 2.5~GeV were reported from $p(K^-,K^+)$ reaction from the measurement of the missing mass of $K^+$~\cite{bib14}. 
The experimental situation with $\Omega^{-\ast}$'s is even worse than the $\Xi^\ast$ case, there are very few data for excited states. The main reason for such a scarce dataset in multi strange hyperon domain is mainly due to very low cross section in indirect production with pion or in particular- photon beams. 
Currently only ground state $\Omega^-$ quantum numbers are identified. Recently significant progress is made in lattice QCD calculations of excited baryon 
states~\cite{bib15,bib16} which poses a challange to experiments to map out all predicted states.
 The advantage of baryons containing one or more strange quarks for lattice calculations is that then number of open decay channels is in general smaller than for baryons comprising only the light u and d quarks. Moreover, lattice calculations show that there are many states with strong gluonic content in positive parity sector for all baryons.  The reason why hybrid baryons have not attracted the same attention as hybrid mesons is mainly due to the fact that they lack manifest ``exotic" character. Although it is difficult to distinguish hybrid baryon states, there is significant theoretical insight to be gained from studying spectra of excited baryons, particularly in a framework that can simultaneously calculate properties of hybrid mesons. Therefore this program will be very much complementary to the GlueX physics program of hybrid mesons.

The proposed experiment  with a beam intensity $10^4 K_L$/sec will result in about $2\times 10^5$~$\Xi^\ast$'s  and $4\times 10^3$~$\Omega^\ast$'s per month.

A similar program for KN scattering is under development at J-PARC with charged kaon beams~\cite{bib17}. The current maximum momentum of secondary beamline of 2~GeV/c is available at the K1.8 beamline. The beam momentum of 2~GeV/c corresponds to $\sqrt{s}$=2.2~GeV in the 
$K^-p$ reaction which is not enough to generate even the first excited $\Xi^\ast$ state predicted in the quark model. However, there are plans to create high energy beamline in the momentum range 5-15~GeV/c to be used with the spectrometer commonly used with the J-PARC P50 experiment which will lead to expected yield of $(3-4)\times10^5$~$\Xi^\ast$'s and $10^3$~$\Omega^\ast$'s per month. 

Statistical power of proposed experiment with $K_L$ beam at JLab will be of the same order  as that in J-PARC with charged kaon beam.

An experimental program with kaon beams will be much richer and allow to perform a complete experiment using polarized target and measuring recoil polarization of hyperons. This studies are under way to find an optimal solution for GlueX setup.

\section{SUMMARY}

In summary we intend to create high intensity $K_L$ beam using photoproduction processes from a secondary Be target.
A flux as high as $10^4 K_L$/sec could be achieved. Momenta of $K_L$ beam particles will be measured with time of flight. The flux of kaon beam will be measured through partial detection of $\pi^+\pi^-$ decay products from their decay to $\pi^+\pi^-\pi^0$ by exploiting similar procedure used by LASS experiment at 
SLAC~\cite{bib11}.
Besides using unpolarized $LH_2$ target currently installed in GlueX experiment additional studies are needed to find the optimal choice of polarized targets. This proposal will allow to measure $KN$ scattering with different final states including production of strange and multi strange baryons with unprecedented statistical precision to test QCD in non perturbative domain. It has a potential to distinguish between  different quark models and test lattice QCD predictions for excited baryon states with strong hybrid content.

\section{ACKNOWLEDGMENTS}
I am indebted to Ilya Larin  for his help in preparing this talk. My research is supported by the U.S. Department of Energy under  grant DE-FG02-96ER40960.


\nocite{*}
\bibliographystyle{aipnum-cp}%
\bibliography{klong}%

\end{document}